# Synergistic effect of Si-hydroxyapatite coating and VEGF adsorption on Ti6Al4V-ELI scaffolds for bone regeneration in an osteoporotic bone environment

I. Izquierdo-Barba[1,2,δ], L. Santos-Ruiz[2,3,4,δ], J. Becerra,[2,3,4] M. J. Feito[5], D. Fernández-Villa[5], M. C. Serrano[6], I. Díaz-Güemes[7], B. Fernández-Tomé[7], S. Enciso[7], F. M. Sánchez-Margallo[7], D. Monopoli[8], H. Afonso[8], M. T. Portolés[5\*], D. Arcos[1,2\*] M. Vallet-Regí[1,2\*]

1. Dpto. de Química en Ciencias Farmacéuticas, Facultad de Farmacia, Universidad Complutense de Madrid, Instituto de Investigación Sanitaria Hospital 12 de Octubre i+12, Plaza Ramón y Cajal s/n, 28040 Madrid, Spain

2. CIBER de Bioingeniería Biomateriales y Nanomedicina (CIBER-BBN), Spain

3. Dpto. de Biología Celular, Genética y Fisiología, Instituto de Investigación Biomédica de Málaga (IBIMA), Universidad de Málaga, Spain

4. Andalusian Centre for Nanomedicine and Biotechnology (BIONAND), c/ Severo Ochoa 35, 29590 Campanillas-Málaga, Spain

5. Dpto. de Bioquímica y Biología Molecular, Facultad de Ciencias Químicas, Universidad Complutense de Madrid, Instituto de Investigación Sanitaria del Hospital Clínico San Carlos (IdISSC), Ciudad Universitaria, 28040 Madrid, Spain.

6. Instituto de Ciencia de Materiales de Madrid (ICMM), Consejo Superior de Investigaciones Científicas (CSIC), 28049 Madrid, Spain

7. Centro de Cirugía de Mínima Invasión Jesús Usón, Cáceres, Spain

8. Dpto. Ingeniería Biomédica. Instituto Tecnológico de Canarias, Spain

\* Corresponding authors: portoles@quim.ucm.es (M.T. Portolés), arcosd@ucm.es (D. Arcos) and vallet@ucm.es (M. Vallet-Regí)

δ These authors contributed equally to this work

**Abstract**

The osteogenic and angiogenic responses to metal macroporous scaffolds coated with silicon substituted hydroxyapatite (SiHA) and decorated with vascular endothelial growth factor (VEGF) have been evaluated *in vitro* and *in vivo*. Ti6Al4V-ELI scaffolds were prepared by electron beam melting and subsequently coated with $Ca_{10}(PO_4)_{5.6}(SiO_4)_{0.4}(OH)_{1.6}$ following a dip coating method. *In vitro* studies demonstrated that SiHA stimulates the proliferation of MC3T3-E1 pre-osteoblastic cells, whereas the adsorption of VEGF stimulates the proliferation of $EC_2$ mature endothelial cells. *In vivo* studies were carried out in an osteoporotic sheep model, evidencing that only the simultaneous presence of both components led to a significant increase of new tissue formation in osteoporotic bone.

# 1. Introduction

Osteoporosis is the most prevalent skeletal disorder in humans older than 50, involving a significant impact on fracture prevalence in the elderly [1,2]. This disease entails a decrease in both bone density and quality, leading to fractures that frequently involves disability and even death. [3] Certainly, fractures are the main clinical consequences of osteoporosis, which are often severely comminuted, especially in trabecular bone areas [4]. But this scenario is even more complex when osteoporotic patients undergo severe trauma entailing the loss of bone mass. The mesenchymal stem cells of osteoporotic bone have less capacity to differentiate into osteoblasts, together with a decreased angiogenic capacity at the defect site [5]. In these cases, biomaterials used for bone grafting and augmentation must also cope with delayed bone healing and impair osseointegration [6-9]. In this sense, strategies such as association of antiosteoporotics drugs to bone grafts have been previously considered [10-12].

An alternative to the current substitutive strategies in the treatment of bone defects is the concept of functionalized metallic macroporous scaffolds [13]. These scaffolds must facilitate the osteogenesis and new blood vessels formation within their macroporous structure, while exhibiting optimal mechanical behavior. Both aspects are mandatory for the regeneration of bone defects, particularly in osteoporotic bones.

The surface functionalization of these metal structures with a highly bioactive bioceramic has emerged as a very promising alternative [14,15]. Moreover, these bioceramics can facilitate fixation of growth factors involved in the osteogenic processes [16-19]. Silicon substituted hydroxyapatite (SiHA) has become one of the most attractive bioceramics for use as bone substitute material [20-23] for spinal, orthopedic, periodontal, oral and craniomaxillofacial applications. SiHA presents enhanced bioactivity *in vivo* than hydroxyapatite (HA), showing beneficial effects in the early stages of bone formation [24]. The favorable effects of Si substitution in HA have been explained by considering passive and active mechanisms as material solubility increase, topographical changes, grain size reduction, surface charge modifications and ionic release of Si and Ca, which directly act on bone cells [25-29].

Vascular endothelial growth factor (VEGF) is involved in angiogenesis and vascular homeostasis [30] and plays an essential role for regulating angiogenesis, endothelial cell function and signaling [31]. The angiogenic process accompanies bone regeneration acting as limiting factor for the healing process [32]. More specifically VEGF regulates angiogenesis, maturation of osteoblasts, ossification, and bone turnover. Thus, osteogenesis and vascularization are coupled during bone development and growth [33-36].

In this study, we evaluated the osteogenic and angiogenic responses to bone implants coated with Si-HA and VEGF for bone defects in an osteoporotic sheep model. Macroporous scaffolds of Ti6Al4V-ELI fabricated by electron beam melting were used as supporting structures and substrates to be coated with SiHA and incubated with VEGF to carry out the adsorption of this growth factor on their surface with a minimal desorption [19]. Ti6Al4V-ELI implants provide strong scaffolding to the bone while exhibiting porosities higher than 50% in volume. Consequently, it would be highly desirable if they exhibited osteogenic capabilities to enhance bone ingrowth within the macroporous structure.

We hypothesized that the presence of immobilized VEGF together with Si-HA would result in a better angiogenic response and stimulation of new bone formation from both, inner site and peripheral area of the bone defect. To test this hypothesis, we evaluated the *in vitro* response of endothelial and pre-osteoblastic cells to the above mentioned Ti6Al4V-ELI scaffolds, as a previous study before implanting them into a cavitary defect of osteoporotic sheep, then evaluating the new bone and blood vessels formation.

## 2. Materials and Methods

*2.1 Scaffolds preparation and coating*

Macroporous structures of Ti6Al4V ELI were prepared by electron beam melting. Cylinders of 1 cm in diameter and 1 cm in height were designed containing a 3D interconnected macroporosity (supporting information Fig S1). The cylindrical structures exhibit pores of 2 mm in diameter and wall thickness around 700 – 800 µm, thus resulting in highly porous structures with large free volume to allow bone ingrowth. The Ti6Al4V – ELI scaffolds so manufactured were coated with Si-HA of nominal composition $Ca_{10}(PO_4)_{5.6}(SiO_4)_{0.4}(OH)_{1.6}$, following a dip-coating process. Aqueous sols were prepared by hydrolyzing 2.58 mL and 0.136 mL of triethyl phosphite $P(OCH_2CH_3)_3$ (TIP) and tetraethyl orthosilicate $Si(OCH_2CH_3)_4$ (TEOS), respectively, in 1.04 mL of $H_2O$. The solution was stirred at 200 rpm for 24 hours to hydrolyze both alkoxides. This solution was subsequently poured to a second solution of 1.21 g of a non-ionic surfactant Pluronic F127 ($EO_{106}PO_{70}EO_{106}$) in 13.1 g of ethanol. After 30 min under stirring, 6.25 mL of 4 M $Ca(NO_3)_2 \cdot 4H_2O$ solution were added, thus keeping Ca/P+Si molar ratio to 1.67 as corresponds to $Ca_{10}(PO_4)_{5.6}(SiO_4)_{0.4}(OH)_{1.6}$. The mixed sol was stirred for 15 min and subsequently aged at 60ºC for 24 hours. After that, ethanol was added until doubling the volume. The sol was deposited by the dip-coating method at room temperature with a withdrawal rate of 1,000 µm/s, dried in air for 1 hour at room temperature and subsequently annealed at 550ºC for 10 minutes under air atmosphere to remove the surfactant and to produce the SiHA phase. With the aim of obtaining thicker coatings, this procedure was repeated three times.

*2.2 Immobilization of VEGF on Ti6Al4V structures*

Non-coated Ti6Al4V (Ti) and SiHA-coated Ti6Al4V scaffolds (Ti@SiHA) were introduced into 24 well culture plates (CULTEK S. L. U., Madrid, Spain) and sterilized under ultraviolet light during 1 hour for each side in a sterile environment. Adsorption and immobilization of Vascular Endothelial Growth Factor (VEGF) on scaffold surfaces was carried out through non-covalent binding by incubation of each scaffold with 5 μg/mL of VEGF-A (VEGF-121, 583204, Biolegend, San Diego, CA, USA) in phosphate buffered saline (PBS, Sigma-Aldrich, St Louis, MO, USA) at 4 ºC for various times. Cylinders of 1 cm in diameter and 0.5 cm in height were used for the *in vitro* studies and coated with 500 μL of VEGF solution. For the *in vivo* studies, cylinders of 1 cm in diameter and 1 cm in height were coated with 1 ml of VEGF solution. The samples were stored at 4ºC for 24 h before the *in vitro* and *in vivo* studies. After 0, 0.5 and 24 hours of incubation, the concentration of desorbed VEGF in the supernatants was analyzed by using an enzyme linked immunosorbent assay (ELISA, Cloud-Clone Corp, USA). The adsorbed VEGF amount was indirectly calculated as the difference between the VEGF levels at the initial time and after each incubation time. The standard curve was carried out according to the manufacturer's instructions. The sensitivity of these assays was less than 6.1 pg/mL. In agreement with previous studies [19], a minimal desorption of immobilized VEGF was observed.

Four groups of samples were used in these *in vitro* and *in vivo* studies:

- Ti6Al4V-ELI scaffolds (Ti)
- Ti6Al4V-ELI scaffolds with adsorbed VEGF (Ti-VEGF)
- Ti6Al4V-ELI scaffolds coated with Si-HA (Ti@SiHA)
- Ti6Al4V-ELI scaffolds coated with Si-HA and adsorbed VEGF (Ti@SiHA-VEGF)

*2.3 Physico-chemical characterization*

The scaffolds were characterized by X-ray diffraction (XRD), scanning electron microscopy (SEM) and Fourier transform infrared spectroscopy (FTIR). XRD patterns were collected directly from the coated scaffolds using the grazing incidence technique with a Philips X'Pert diffractometer ($K_\alpha$Cu radiation, $\lambda =1,5418$ Å). Scanning electron microscopy was carried out with a JEOL 6400 Microscope-Oxford Pentafet super ATW microscope equipped with a LINK "Pentafet" detector for EDX analyses. Fourier transform infrared (FTIR) spectroscopy was carried out with a Nicolet Magma IR 550 spectrometer. The samples were prepared by scratching the coatings deposited on the Ti6Al4V scaffolds. The spectra were collected using the attenuated total reflectance (ATR) technique with a Golden Gate ATR device.

*2.4 Culture of endothelial cells on Ti6Al4V structures*

Mature endothelial cells $EC_2$ obtained from porcine peripheral blood (see supporting information) were seeded on the scaffolds at a density of $3 \times 10^5$ cells/mL in EGM-2 (Lonza, Walkersville, MD, USA). EGM-2 without VEGF was used for the culture of $EC_2$ on these scaffolds in order to evaluate the effects of the VEGF immobilized on scaffold surface. All the samples were incubated under 5% $CO_2$ atmosphere at 37 ºC for 3 days.

*2.5 Culture of MC3T3-E1 pre-osteoblasts on Ti6Al4V structures*

Murine pre-osteoblastic MC3T3-E1 cells (subclone 4, CRL-2593, ATCC, Manassas, VA, USA) were seeded on scaffolds at a density of $6 \times 10^5$ cells/mL in Dulbecco's Modified Eagle Medium (DMEM, Sigma Chemical Company, St. Louis, MO, USA) supplemented with 10% fetal bovine serum (FBS, Gibco, BRL), 1 mM L-glutamine (BioWhittaker Europe, Belgium), penicillin (200 µg/mL, BioWhittaker Europe, Belgium), and streptomycin (200 µg/mL, BioWhittaker Europe, Belgium). All the samples were incubated under 5% $CO_2$ atmosphere at 37 ºC for 3 days.

*2.6 Cell proliferation assay: CCK-8*

Proliferation of both endothelial cells ($EC_2$) and MC3T3 pre-osteoblasts was measured using the Cell Counting Kit-8 (CCK-8) protocol (Sigma-Aldrich, St Louis, MO, USA). After incubation for 3-4 hours under 5% $CO_2$ atmosphere at 37 ºC, three samples of 100 µL of each well were collected into 96 well culture plates (Nunc Brand, Rochester, NY, USA) to have triplicates of each initial sample and absorbance was measured at 450 nm.

*2.7 Scanning electron microscopy (SEM) studies*

After CCK-8 protocol, the samples were used for SEM analysis because WST-8 does not damage living cells. Cells attached to the scaffolds were fixed by incubation with 2.5% glutaraldehyde (Merck KGaA, Darmstadt, Germany) for 1 h at 4 ºC. Then, successive dehydration steps were carried out by slow water replacement, using a series of ethanol solutions (30%, 50%, 70%, 80% and 90%) for 10 min and final dehydration in absolute ethanol for 10 min, allowing the samples to dry at room temperature. Afterwards, the pieces were mounted on stubs and coated in vacuum with gold–palladium. Samples were then examined with a JEOL JSM-6400 scanning electron microscope (Centro Nacional de Microscopía Electrónica, Madrid, Spain).

*2.8 In vivo studies in osteoporotic sheep model*

This study was approved by our Institutional Ethical Committee following the guidelines of the current normative (Directive 2010/63/EU of the European Parliament and of the Council of September 22, 2010, on the protection of animals used for scientific purposes).

*Induction of osteoporotic model*

Six 4-year-old female Merino sheep (mean preoperative weight of 43.78 ± 5.9 Kg) were included in the study, and all of them were operated on to place with the four implants described above at distinct locations. To reproduce similar conditions as osteoporosis in humans, six months before the implantation all sheep underwent, under sterile conditions and general anaesthesia induced by propofol (4 mg/kg) and maintained by isoflurane (1.5%), a laparoscopic bilateral ovariectomy (see supporting information for further details). At the same time, a low-calcium diet (0.5%) and corticosteroids administration (500 mg methylprednisolone via intramuscular injection every 3 weeks) were implemented until the end of the study.

*Implantation surgical procedure*

Six months after the ovariectomy, the biomaterials were blindly implanted in the sheep under aseptic conditions and the same anaesthetic protocol previously described. Six cylindrical defects (10x13mm) were created in each sheep by drilling the cancellous bone of the proximal tibia epiphysis, medial epicondyle of the femur and greater tuberosity of the humerus, under continuous irrigation with cold sterile saline, as described by Nuss *et al.* [37]. The sample size for each type of scaffold was $n = 6$. Two defects in each sheep were left empty as control. Once the biomaterials were randomly implanted, the muscular and subcutaneous tissue was approximated with absorbable monofilament suture and the skin with absorbable braided suture (Figure S2 in supporting information). Postoperative analgesia was maintained with buprenorphine (0.01mg/kg/8h/3 days) and meloxicam (0.4mg/kg/24h/7 days). Ceftiofur (1mg/kg/24h) was administered for 7 days as prophylactic antibiotherapy. The health condition of all animals was checked daily along the whole study by an accredited veterinarian. Immediately after the surgical procedure and before the sample removal, a computed tomography (CT) scan was performed (Figure S3 in supporting information). The sample size for each type of scaffold was $n = 6$.

*2.9 Histological processing*

The bone segments containing the defect were dissected out and fixed by immersion in 10% neutral buffered formalin (4% formaldehyde) for 7 days. Bone segments were divided into 4 mm-thick slices with an EXAKT 300 CP band-saw, and post-fixed in the same fixative for another 4 days. Bone slices were dehydrated by immersion in a graded ethanol series and embedded in Technovit 7210 VLC resin, which was light-polymerized. Polymerized resin blocks were glued to microscope slides and sectioned into 0.4 mm-thick sections with a diamond saw. These sections were polished with EXAKT 400 GRINDING SYSTEM until obtaining 50 µm-thick histological sections.

Sections were stained with either von Kossa's or Masson-Goldner's Trichrome Stainings. For von Kossa's staining, sections were incubated in 1% silver nitrate aqueous solution under ultraviolet light for 20 minutes and rinsed thoroughly for five times in bi-distilled water. Unreacted silver was removed by washing with 5% sodium thiosulfate aqueous solution, followed by new washes in bi-distilled water. Nuclei were counterstained with Nuclear Fast Red. Masson-Goldner's Trichrome Staining was performed as described by Goldschlager [38]. Histomorphometric quantification of bone ingrowth and blood vessels density was carried out by a blind reviewer (LSR) using the slides obtained from the half height section of the cylindrical scaffolds. For the blood vessels quantification, we worked at higher magnification, taking 21 images from each histological slice and scanning the defect in both vertical and horizontal directions and crossing the center.

*2.10 mage acquisition and analysis*

Images were obtained with an Olympus VS120 photo-microscope and analysed with ImageJ to calculate defect area and bone area in each section, as well as number of blood vessels. Data were plotted and statistically analysed with GraphPad Prism7.

*2.11. Statistics*

*In vitro* cell culture data were expressed as means ± standard deviations of a representative of three experiments carried out in triplicate. Statistical analysis was performed using the Statistical Package for the Social Sciences (SPSS) version 19 software. One-way ANOVA followed by Scheffé *post hoc* test were used to evaluate differences among groups for *in vitro* results and two-way ANOVA to compare the effects of SiHA and VEGF for histological data. In all the statistical evaluations, $p < 0.05$ was considered as statistically significant.

**3. Results**

*3.1 Characterization of Si-HA coatings on Ti6Al4V-ELI scaffolds*

Figure 1.a shows the XRD patterns of the coatings when the sols are aged for 6, 8 and 24 hours. The XRD patterns show the variation of the crystalline phases of the Si-HA coatings as a function of ageing time. The pattern for the sol aged for 6 hours (HASi$_{6H}$) shows diffraction maxima corresponding to a calcium carbonate (CaCO$_3$) as major phase, together with diffraction maxima that can be assigned to an apatite-like phase (HA). The XRD pattern for the sol aged for 8 hours (HASi8$_H$) shows that the maxima corresponding to an apatite phase (HA) increase in intensity respect to the maxima

assignable to $CaCO_3$. Finally, the XRD pattern corresponding to ageing times of 24 hours ($HASi_{24H}$) only shows the diffraction maxima corresponding to an apatite-like phase together with several diffraction maxima assigned to the $\alpha$-Ti phase of the substrate.

Figure 2.b show the FTIR spectra for the SiHA coatings aged for different periods. The spectrum of the coating deposited after 6 hours of ageing ($HASi_{6H}$) shows intense adsorption bands corresponding to the vibration of carbonate ions (1407 and 870 $cm^{-1}$). A broad vibration band (medium intensity) corresponding to low crystalline phosphates at 1050 $cm^{-1}$ and a singlet signal at 520 $cm^{-1}$ characteristics of amorphous phosphate are also observed. The coatings deposited after 8 hours of ageing ($HASi_{8H}$) point out a decrease of carbonate content, whereas the absorption bands for phosphates are more intense and indicate more crystallinity as can be deduced from the doublet signal corresponding to the bending mode of O-P-O bonds in crystalline environments. Finally, coatings deposited from sols aged for 24 hours ($HASi_{24H}$) show the absorption bands corresponding to the bending vibration in crystalline phosphates (doublet at 603 and 567 $cm^{-1}$), P-O stretching bands (1089, 1054, and 961 $cm^{-1}$) and the band corresponding to the librational mode of hydroxyl groups in hydroxyapatite. Besides, the carbonate band at 1400 $cm^{-1}$ significantly decreases, in agreement with the absence of calcite determined by XRD.

Figure 2c shows the SEM images and the EDX spectra of the coatings aged for different times. Sols aged for 6 hours ($HASi_{6H}$) led to heterogeneous coatings exhibiting nodules. These heterogeneities are not observed in $HASi_{8H}$ coatings; however these coatings show cracks, which introduce certain degree of heterogeneity on the surface. Finally, sols aged for 24 hours ($HASi_{24H}$) led to crack-free homogeneous coatings made of silicon substituted apatite as indicated by the EDX spectra (insets in figure 2c).

*3.2 Determination of VEGF adsorption and cell culture tests*

VEGF adsorption on Ti and Ti@SiHA scaffolds reaches over 90% after 30 minutes in both cases ($p < 0.005$) (Figure 2). At this short time, VEGF adsorption on Ti@SiHA is significantly higher than on Ti scaffolds ($p < 0.005$), demonstrating that SiHA coating accelerates VEGF adsorption on these structures. After 24 hours, the adsorbed VEGF percentages were the same in both non-coated Ti and Ti@SiHA scaffolds.

Endothelial cell proliferation is significantly higher on Ti-VEGF and Ti@SiHA-VEGF scaffolds compared to scaffolds without VEGF ($p < 0.005$) (Figure 3.a). The best proliferation results are obtained with endothelial cells cultured on non-coated scaffolds after adsorption of VEGF (Ti-VEGF samples). SEM observations suggest that endothelial cells proliferate on all the scaffolds and cover the surface of both coated and non-coated scaffolds (Figure 3.b).

The *in vitro* response of bone cells to all these structures was evaluated by culturing MC3T3 pre-osteoblasts. Figure 4.a shows that the scaffolds coated with SiHA (Ti@SiHA and Ti@SiHA-VEGF) exhibit a significant increase of pre-osteoblasts proliferation ($p < 0.005$). The higher pre-osteoblasts proliferation was obtained when pre-osteoblasts are cultured on Ti@SiHA-VEGF, although there is not statistical significance when compared with Ti@SiHA scaffolds. Regarding non-coated scaffolds, a significant increase of pre-osteoblast proliferation ($p < 0.05$) was observed in Ti-VEGF respect to Ti.

SEM studies were also carried out with MC3T3 pre-osteoblasts cultured on scaffolds with or without adsorbed VEGF. Pre-osteoblast proliferate on all these materials and completely cover the surface of both coated and non-coated scaffolds (Figure 4.b). More granularity is observed on SiHA-coated samples compared to non-coated ones, probably due to the higher pre-osteoblast proliferation on coated scaffolds.

*3.3 Histological evaluation and histomorphometric quantification of bone ingrowth and blood vessels*

Figure 5 shows the histological sections after 12 weeks of implantation in an osteoporotic sheep animal model. Empty control defects did not get filled with newly-formed bone, although some bone growth was observed at the periphery of the defect, as part of the natural healing reaction. Similar peripheral growth was found in groups implanted with Ti and Ti-VEGF. In these groups, there was also some growth next to the titanium bars closer to the periphery. Groups implanted with Ti@SiHA and Ti@SiHA-VEGF also presented this type of growth but, interestingly, bone was frequently found on the inner struts (black arrows in figure 5.d). In some cases, this bone is connected to the peripheral bone (red arrow in figure 5.d). Quantification of newly-formed bone revealed the extent of ossification to be significantly higher in Ti@SiHA-VEGF as compared to the rest of the groups when considering both, the whole defect and the central third of the defect (Figure 6).

Figure 7 shows histological sections of Ti@SiHA and Ti@SiHA-VEGF (Masson-Goldner's Trichrome Stainings) after 12 weeks of implantation. In the case of the scaffolds with VEGF adsorbed, the images evidence thicker trabeculae and a higher number of blood vessel. The quantitative study (Figure 8) showed a significantly higher number of vessels in Ti@SiHA-VEGF compared with the same scaffold without VEGF (Ti@SiHA). In order to test the uniformity of blood vessels formation, we also quantified the number of vessels formed in the peripheral region and central region of Ti@SiHA and Ti@SiHA-VEGF scaffolds. No significant differences were found respect to the region observed.

**Discussion**

Metallic macroporous scaffolds are a promising alternative to the current substitutive strategies in the treatment of bone defects. These scaffolds facilitate the osteogenesis and new blood vessels formation within their macroporous structure, while exhibiting optimal mechanical behavior. Both aspects are mandatory for bone regeneration, particularly in osteoporotic bones where the implant integration with the hosting bone is seriously affected due to the low bone formation rate in the peri-implant region [40]. These implants provide strong scaffolding to the bone with porosities higher than 50% in volume. Consequently, it would be highly desirable if they exhibited osteogenic capabilities to enhance bone ingrowth within the macroporous structure.

In this work we have achieved this aim by coating microporous Ti6Al4V-ELI scaffolds with silicon substituted hydroxyapatite and subsequent adsorption of VEGF. Previously, different synthesis parameters were strictly controlled to prepare homogeneous and stable coatings. Among them, the ageing time of the sol is a determining parameter in dip coating processes. Ageing times below 24 hours were not enough to obtain crack-free SiHA coatings, resulting in a heterogeneous deposition of material composed by calcium carbonate and calcium phosphate as secondary phase. On the contrary, ageing periods of 24 hours followed by three dipping-withdrawal cycles allow for the complete hydrolysis of phosphate and silicon precursors, so that a unique SiHA phase with Ca/P+Si ratio of 1.67 is obtained. SiHA coating so obtained facilitates VEGF adsorption on these structures. VEGF is adsorbed very efficiently after 30 min in contact with Ti@SiHA scaffolds, although uncoated Ti scaffolds also exhibited high and stable non-covalent adsorption capability for this growth factor. After 24 hours both Ti-VEGF and Ti@SiHA-VEGF showed the same levels of VEGF adsorption close to 100%. Concerning the release kinetics of the VEGF from the scaffold, in preliminary studies we measured *in vitro* by ELISA the VEGF spontaneous desorption in PBS from disks of nanocrystalline and crystalline hydroxyapatites with different Si proportions for different times after VEGF adsorption [19]. Very low levels (lower than 2 ng/ml) of desorbed VEGF were observed 96 hours after the initial time in all the cases, thus indicating the effective immobilization of VEGF.

Although no differences could be detected by SEM between coated and non-coated scaffolds, VEGF seems to improve the endothelial cell adhesion and proliferation on the surface of all these structures. This fact is observed in both series containing VEGF, pointing out the relevance of this growth factor as an important mediator of endothelial cell proliferation and survival [39]. Besides, endothelial cells proliferation is significantly higher in Ti-VEGF compared with Ti@SiHA-VEGF. This result would point out that SiHA coating could decrease the stimulatory proliferation effect elicited by VEGF in $EC_2$ cells.

The scaffolds coated with SiHA (Ti@SiHA and Ti@SiHA-VEGF) exhibit a significant increase of pre-osteoblast proliferation. These results highlight the relevance of this coating for bone cells and agree with the specific characteristics of silicon substituted hydroxyapatite. It is well known that silicon substitution of the hydroxyapatite potentiates pre-osteoblast proliferation and maturation by different established biochemical mechanisms [41]. Besides, a significant increase of pre-osteoblast proliferation ($p < 0.05$) is observed in Ti-VEGF respect to Ti scaffolds, suggesting the effect of VEGF adsorption on Ti6Al4V scaffolds is also remarkable on pre-osteoblasts. On Ti@SiHA scaffolds, the VEGF effect is not significant probably due to the beneficial effect of the coating, which masks the action of this growth factor.

All these *in vitro* studies demonstrate that Ti-VEGF scaffolds improve the proliferation of endothelial cells while pre-osteoblasts prefer to grow on Ti@SiHA-VEGF structures, accordingly to the characteristics and specific functions of the endothelium and bone tissue, respectively. On the other hand, the SEM images of endothelial cells and MC3T3 pre-osteoblasts (Figures 3 and 4, respectively) cultured on the different scaffolds with or without adsorbed VEGF, evidence that these two cell types proliferate on all the scaffolds and cover their surface forming a perfect monolayer which makes difficult to distinguish the cells on the surface. It is important to take into account that the formation of an endothelial cell monolayer plays a key role in the development of tissue engineered vascular grafts and facilitate the organ functions.

In order to evaluate the osteogenic capability of the scaffolds in conditions simulating osteoporosis in humans, we have implanted the four series in an osteoporotic sheep model. This ovine model, combining ovariectomy with a low calcium diet and corticosteroid administration, has been previously described in the literature [42,43]. One limitation of our study was that we could not achieve a dose of calcium of 0.15-0.25% as described by other authors [43,44] due to technical limitations, although it was lower than that found in the normal diet. Nevertheless, an osteoporotic model has been described as well with a combination of ovariectomy and glucocorticoids administration alone determining that it constitutes a relevant preclinical model for orthopaedic implant and biomaterial research [45], being the best effects for osteoporosis induction obtained using ovariectomised sheep with methylprednisolone injections [46]. Even more, Stadelmann *et al.* [47] described an osteoporotic model with ovariectomized sheep after 6 months with no restriction in calcium and no corticosteroids administration. The authors reported a reduction of 30% in bone volume fraction, 11% in trabecular thickness and 19% in trabecular number, as well as a 14% increase in trabecular separation, although the structural model index was not significantly affected. On the other hand, the corticosteroid protocol used was that described by Klopfenstein *et al.* [42] for osteoporotic sheep models in order to minimize the side effects produced by long-term treatments but maintaining the osteopenia conditions. Additionally, a minimally invasive approach was used for the ovariectomy to decrease discomfort and pain in the animals.

After 12 weeks *in vivo*, clear differences were observed between samples. In the control defect, no bone formation (figure 5.a) or new bone formation restricted to peripheral regions was observed (see supporting information, Figure S4), whereas uncoated Ti scaffolds showed moderate bone ingrowth from peripheral region towards the inner parts. The adsorption of VEGF did not modify the bone in-growth pattern, i.e. bone regeneration also occurs from peripheral region. However, the presence of VEGF seems to stimulate the formation of thicker trabeculae in Ti-VEGF compared to Ti scaffolds.

Coating Ti6Al4V scaffolds with SiHA did not improve the amount of newly formed bone respect to Ti. In the absence of VEGF, trabeculae formed on Ti@SiHA are thin and poorly developed, similarly to those observed for naked Ti. However, the scaffolds with SiHA coatings show some bone formation in the inner struts of Ti@SiHA scaffolds in addition to that originated from the peripheral bone. This bone growth might start in the inner struts or there might be out-of-plane growth from the periphery. This fact would agree with the osteoinductive properties of silicon-substituted hydroxyapatites previously pointed by Patel et al [48,49], although more work, like the quantification of bone growth patterns in 3D, would be needed to identify if bone formation does truly start at the inner struts.

Despite of the positive outcomes associated to the presence of SiHA coatings and VEGF separately (Ti@SiHA and Ti-VEGF scaffolds), only Ti@SiHA-VEGF scaffolds exhibited significantly higher bone formation (50.6 ± 22.3% bone volume) than either Ti@SiHA, Ti-VEGF or Ti scaffolds (19.6 ± 5.6%, 26.7 ± 12.6% and 17.8 ± 13.2 % bone volume, respectively). In addition, the newly formed bone exhibited thicker trabeculae with a more developed vascular system.

These results must be understood in terms of the multiple effects that SiHA coating and VEGF can exert on different cell types, as well as the role of each one plays in osteogenesis and angiogenesis. Coating Ti6Al4V-ELI scaffolds with SiHA and subsequent adsorption of VEGF suggest an interplay between the bioactive coating and the grow factor for bone regeneration. As we could observe in our *in vitro* cell culture tests, SiHA coating stimulates pre-osteoblast proliferation on the scaffolds surface. *In vivo* studies carried out with rabbits demonstrated the capability of silicon substituted hydroxyapatites to promote bone ingrowth and repair, characterized by a dense trabecular morphology [50]. These biological outcomes are sensitive to the silicon content, as it was also observed in an ovine model where the rate and quality of bone apposition was enhanced with Si substitution [48]. This enhanced osteogenic performance is explained in terms of the excellent osteoconductive properties provided by the hydroxyapatite phase, together with the significant up-regulation of osteoblast proliferation and gene expression mediated by soluble silica species [51,52]. Besides, our *in vitro* cell culture tests demonstrate that VEGF adsorbed on Ti and Ti@SiHA scaffolds stimulates the proliferation of endothelial cells. The excellent osteogenic response in this sheep model for osteoporosis could be explained in terms of a previous stimulated angiogenesis within the defect and subsequent recruitment and supporting of osteoprogenitor cells. Although the VEGF dose used in this study was chosen in agreement with recent studies by others [53], this response would be reinforced by the osteoinductive capability of SiHA coated scaffolds, which are able to stimulate new bone formation even in the inner sites of the defect not in contact with the peripheral bone.

## 4. Conclusions

Macroporous Ti6Al4V scaffolds coated with silicon-substituted hydroxyapatite and decorated with VEGF have been prepared and evaluated as bone graft for bone regeneration purposes.

*In vitro* cell culture tests evidence that those scaffolds with adsorbed VEGF stimulate proliferation of endothelial cells on the scaffolds surface, whereas those scaffolds coated with SiHA stimulate proliferation of pre-osteoblasts.

*In vivo* studies on a sheep model for osteoporosis evidence that neither SiHA coating nor VEGF adsorption enhance osteogenesis separately. However, the adsorption of VEGF on SiHA coated scaffolds exhibits a synergistic effect resulting in more ossification, larger trabeculae and higher angiogenesis degree.

**Acknowledgements**

This study was supported by research grants from the Ministerio de Economía y Competitividad (projects MAT2015-64831-R, MAT2016-75611-R AEI/FEDER, UE and BIO2015-66266-R), and Instituto de Salud Carlos III (RD12-0019-0032) M.V.-R. acknowledges funding from the European Research Council (Advanced Grant VERDI; ERC-2015-AdG Proposal 694160). The authors wish to thank the ICTS Centro Nacional de Microscopia Electrónica (Spain) of the Universidad Complutense de Madrid (Spain) for their technical assistance.

**Figure captions**

**Figure 1.** XRD patterns (A), FTIR spectra (B) and SEM images (C) of the coatings deposited on Ti6Al4V scaffolds after 6, 8 and 24 hours of aging. The insets show the EDX spectra obtained from the scaffolds surface during SEM observations

**Figure 2.** Indirect evaluation of VEGF adsorption on Ti (white) and Ti@SiHA (grey) scaffolds. VEGF union percentage at various times was analyzed by measuring free VEGF in the supernatants by ELISA. *Comparison between each material at various times. ɸ Comparison between coated and non-coated scaffolds. Statistical significance: *** $p < 0.005$; ɸɸɸ $p < 0.005$.

**Figure 3.** (a) Proliferation of $EC_2$ endothelial cells cultured on Ti and Ti@SiHA scaffolds without or with adsorbed VEGF. Cell proliferation was analyzed by measuring absorbance at 450 nm after CCK-8 protocol. *Comparison between each material without or with VEGF. ɸ Comparison between coated and non-coated scaffolds. Statistical significance: *** $p < 0.005$; ɸ $p < 0.05$. SEM studies of $EC_2$ endothelial cells cultured on Ti (b), Ti-VEGF (c), Ti@SiHA (d) and Ti@SiHA-VEGF scaffolds (e)

**Figure 4**. (a) Proliferation of MC3T3 pre-osteoblasts cultured on Ti and Ti@SiHA scaffolds without or with adsorbed VEGF. Cell proliferation was analyzed by measuring absorbance at 450 nm after CCK-8 protocol. *Comparison between each material without or with VEGF. ɸ Comparison between coated and non-coated scaffolds. Statistical significance: * $p < 0.05$; ɸɸɸ $p < 0.005$. SEM studies of MC3T3 pre-osteoblasts cultured on Ti (b), Ti-VEGF (c), Ti@SiHA (d) and Ti@SiHA-VEGF scaffolds (e)

**Figure 5**. Optical micrographs from the histological sections after 12 weeks (von Kossa's staining): (a) control defect, (b) Ti scaffold, (c) Ti-VEGF scaffold, (d) Ti@SiHA scaffold. The inset is a magnification of an inner strut with new bone grown on it. (e) Ti@SiHA-VEGF. Arrows indicate bone growth non-derived from the border of the defect.

**Figure 6.** Defect area covered by new bone. (a) new bone in the whole defect and (b) new bone in the central third of the defect. Statistical significance: * $p < 0.05$

**Figure 7.** Optical micrographs from the histological sections after 12 weeks (Masson-Goldner's Trichrome Stainings) obtained from the peripheral regions of (a) Ti@SiHA and (b) Ti@SiHA-VEGF scaffolds. Arrows point to blood vessels

**Figure 8.** Quantitative evaluation of angiogenesis in Ti@SiHA and Ti@SiHA-VEGF scaffolds in the central and peripheral regions of the scaffolds. Statistical significance: * $p < 0.05$.

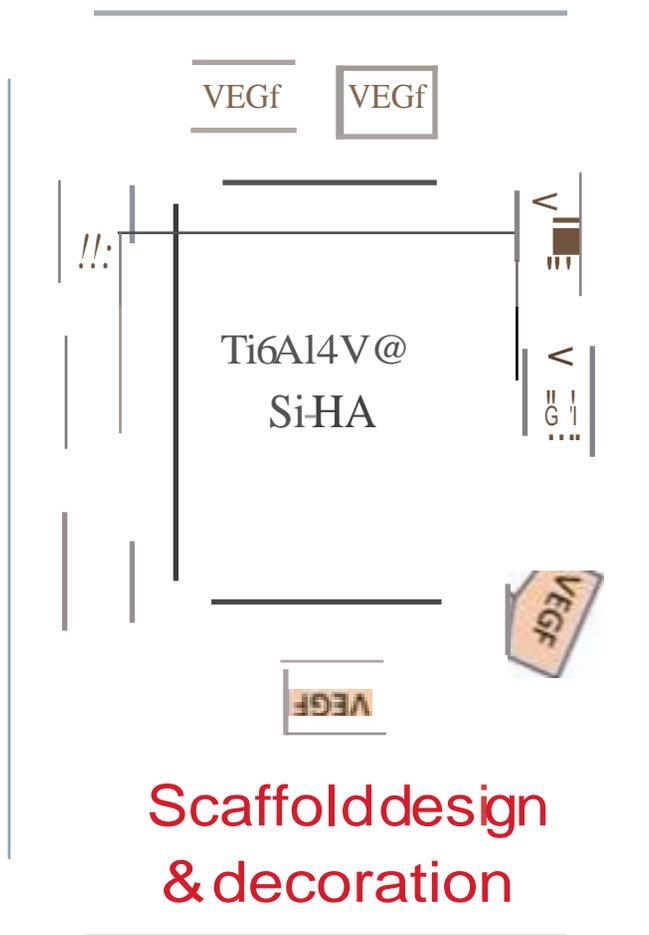
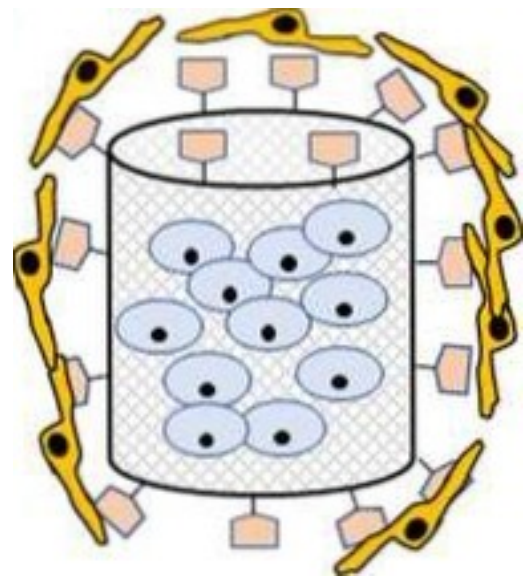
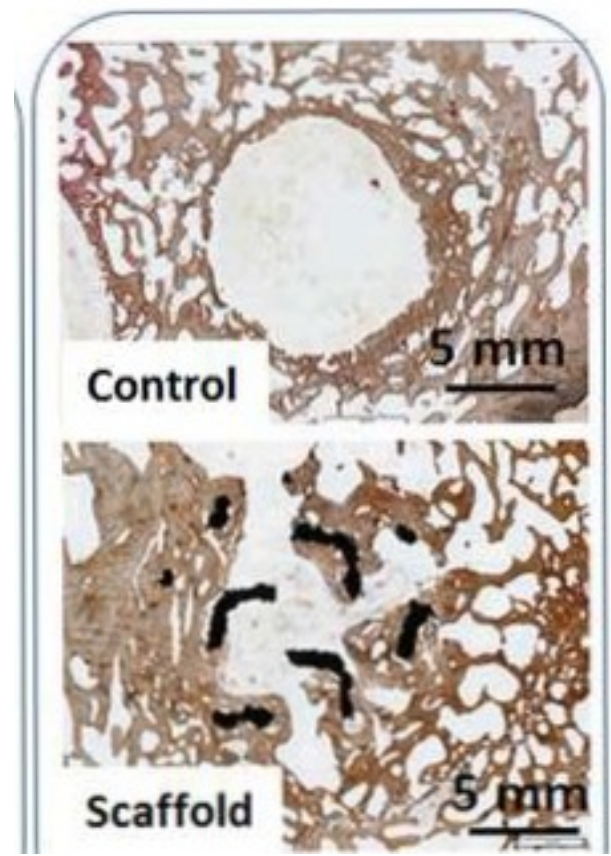

Scaffold design & decoration

G) Pre-osteoblast
Endothelial cell

*In vitro* behaviour

*In vivo* behaviour

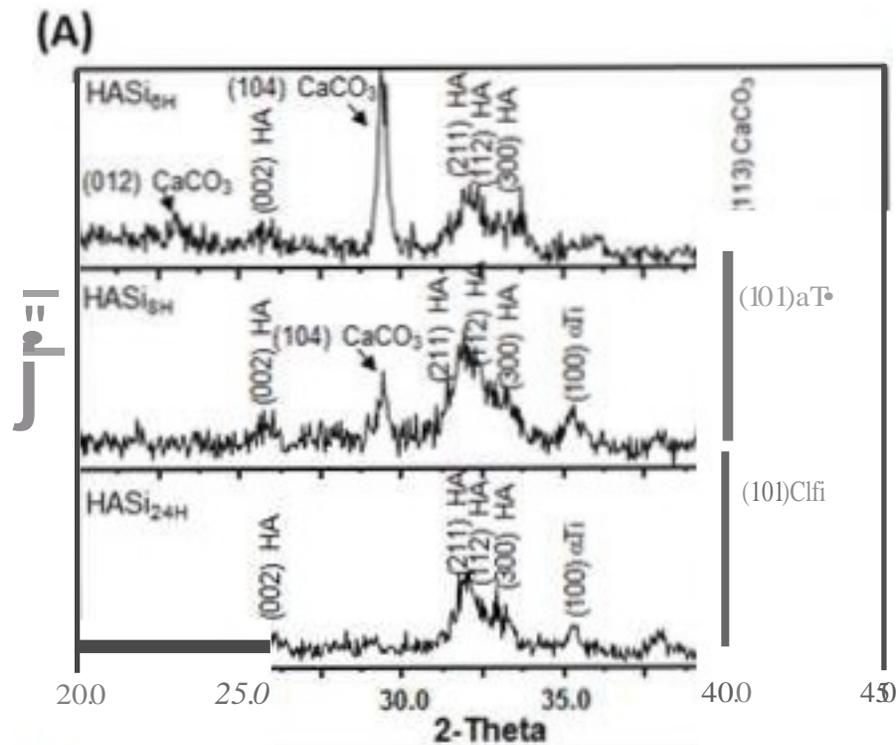
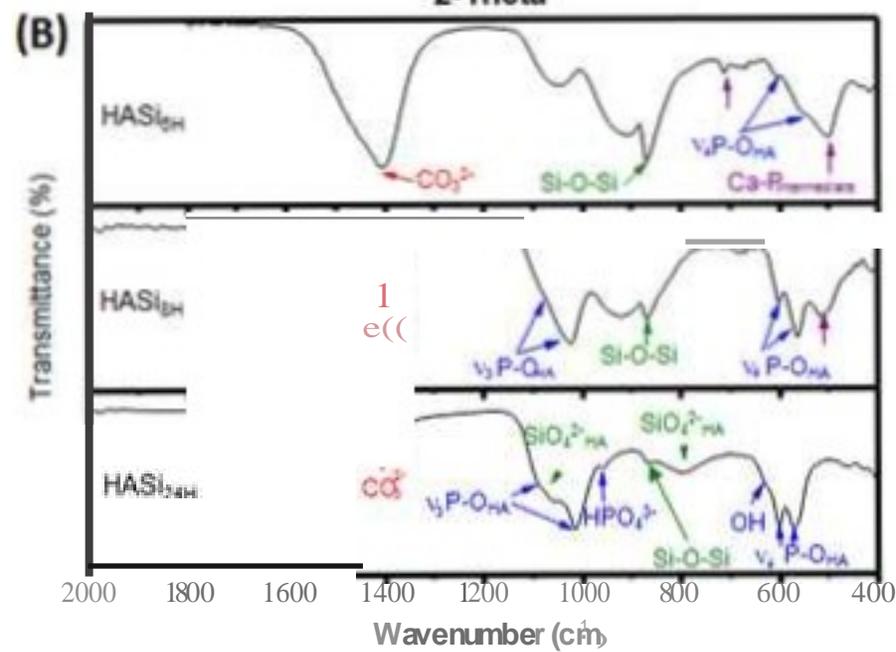
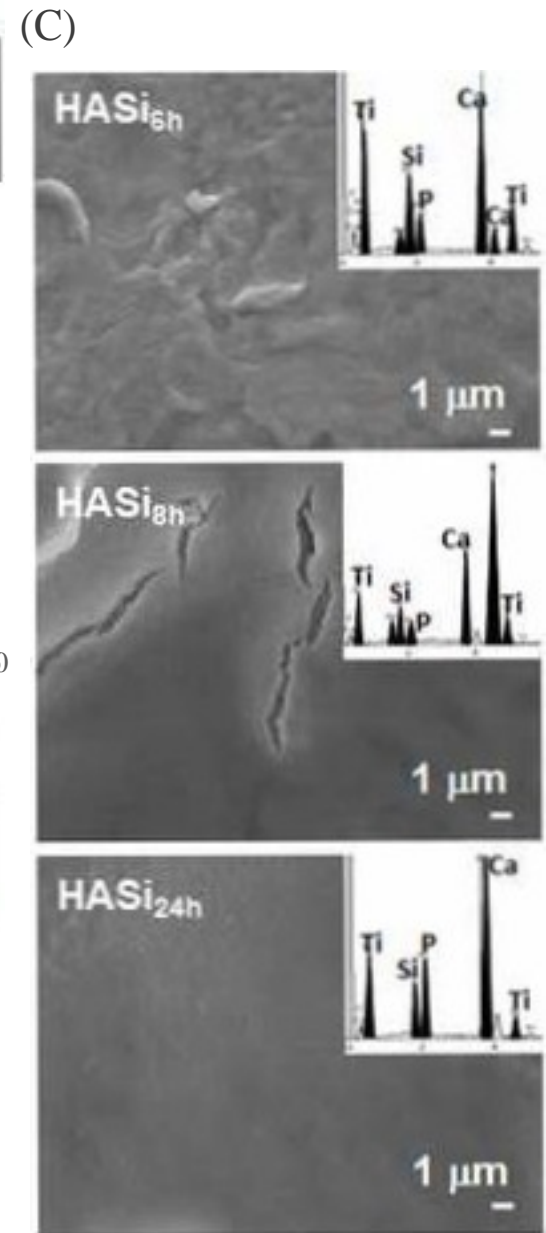

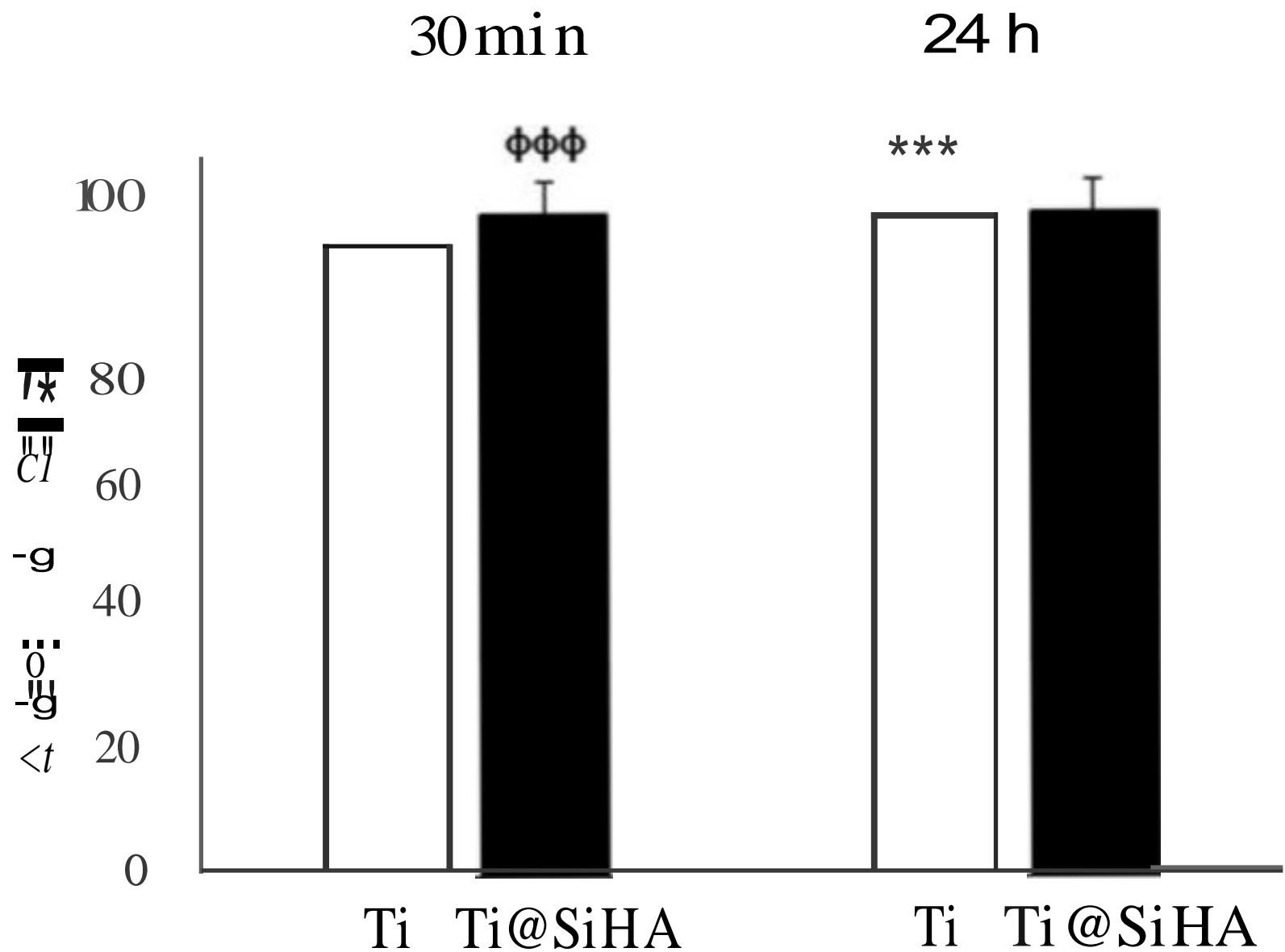

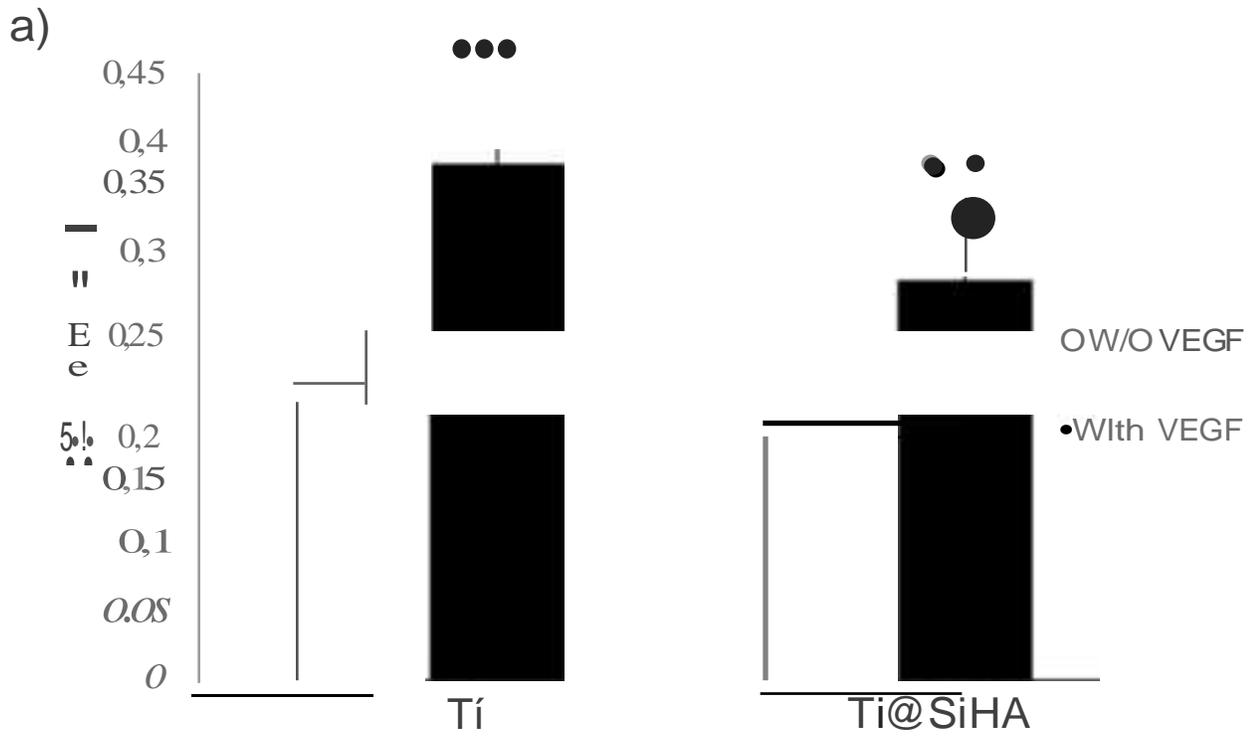
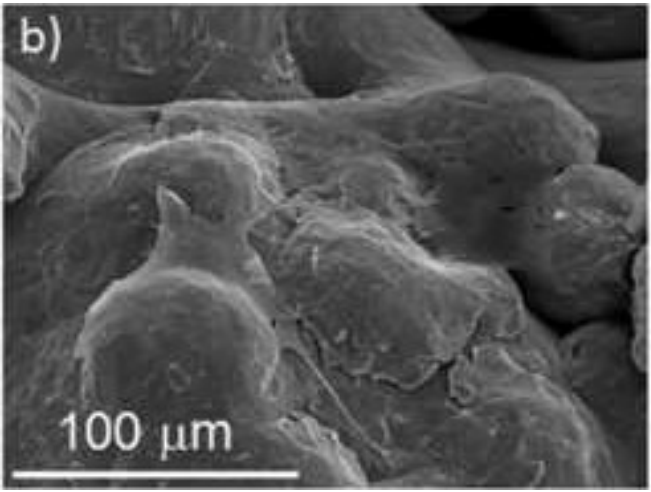
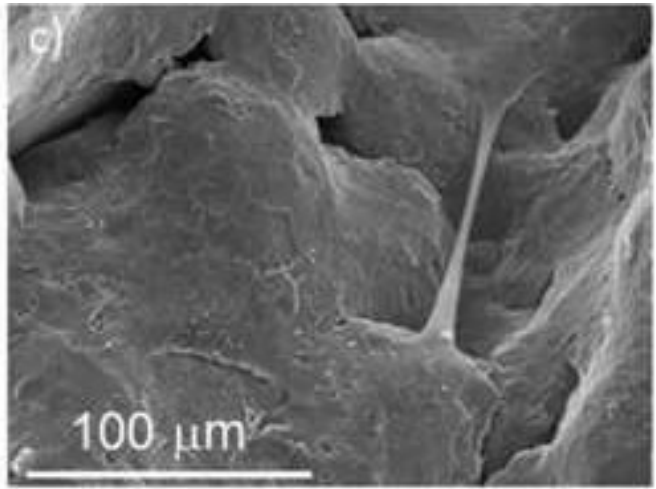
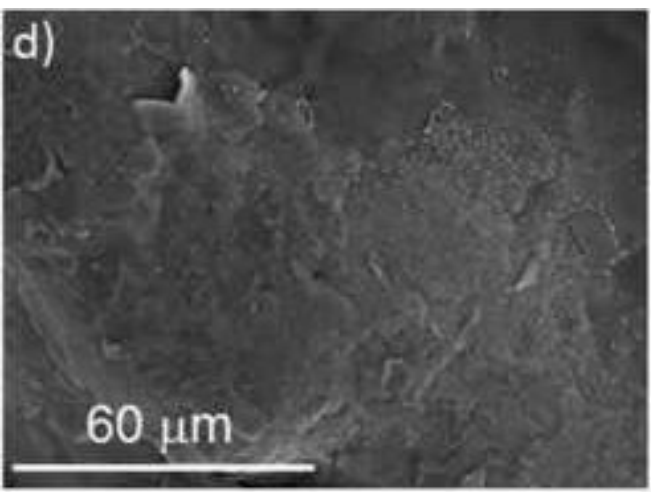
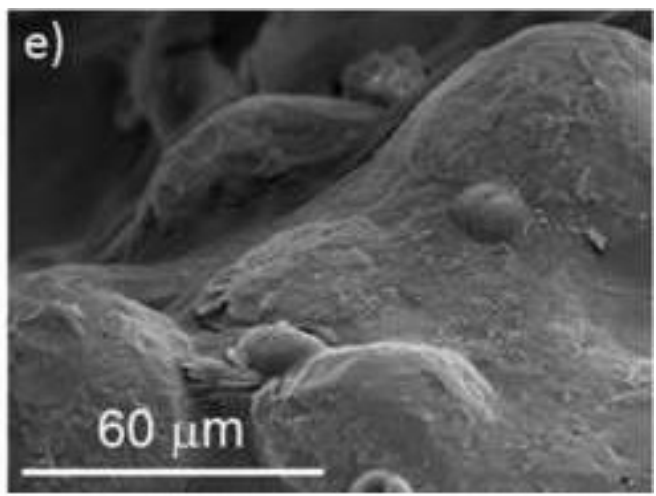

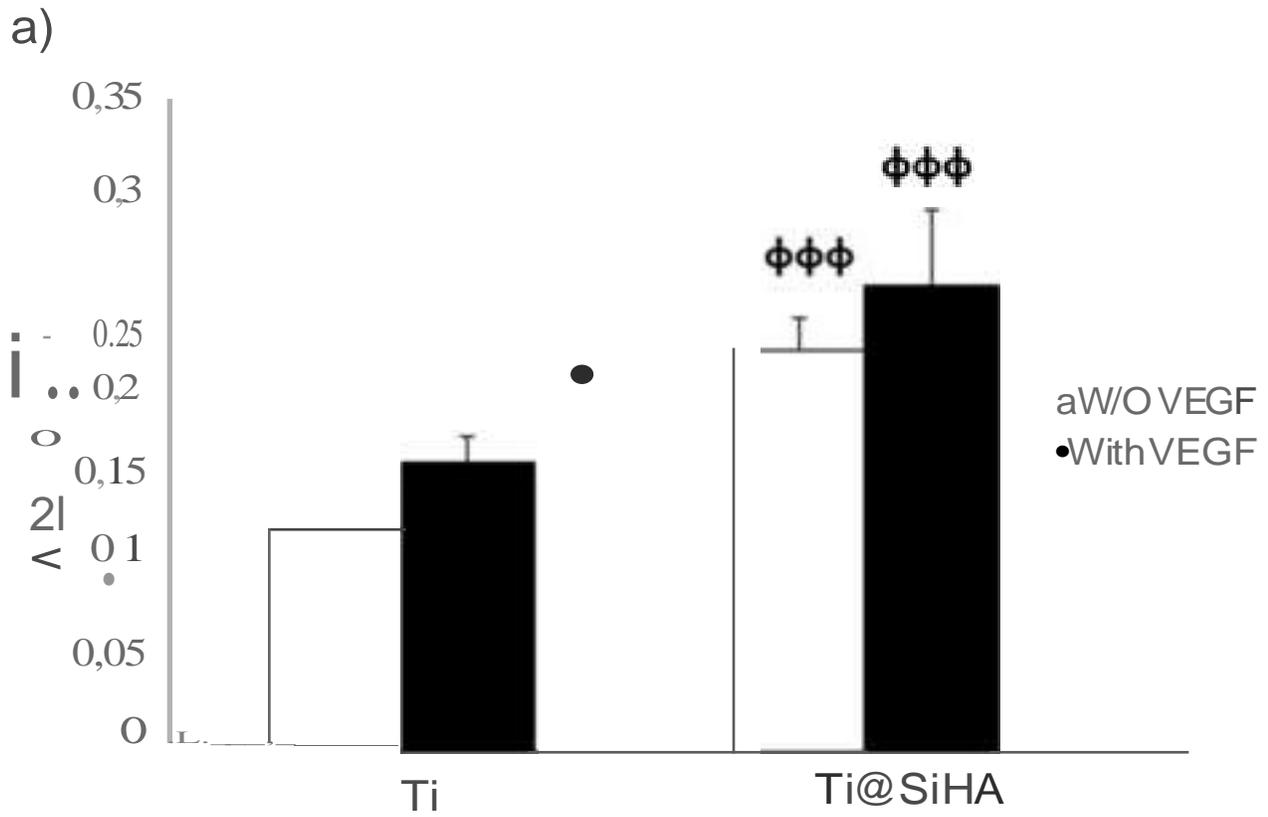
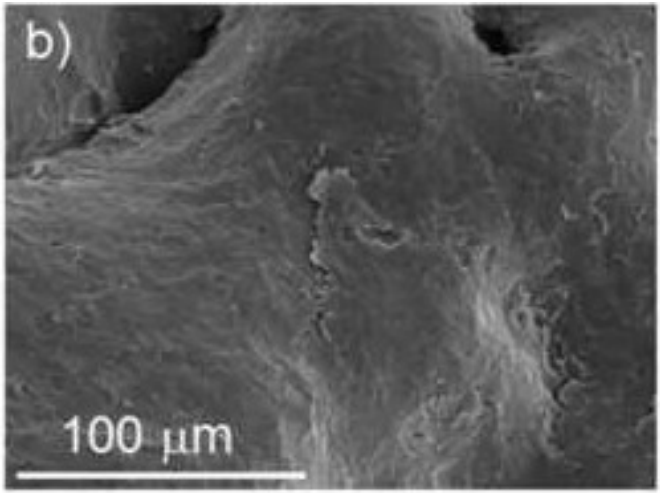
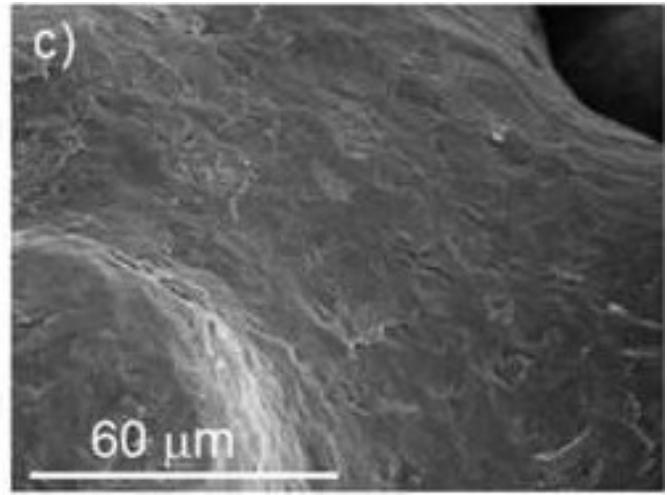
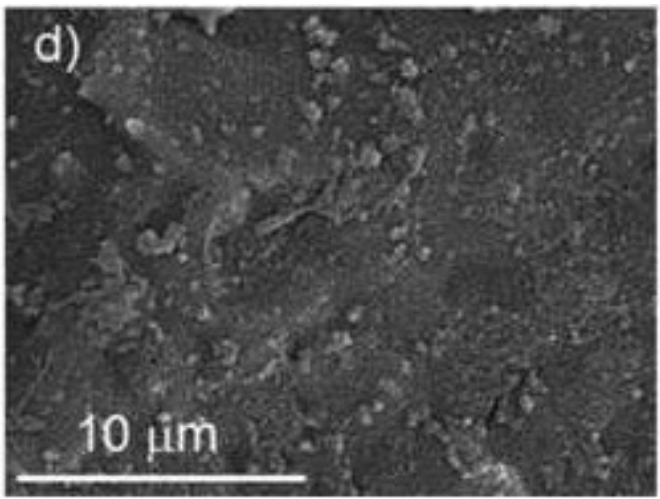
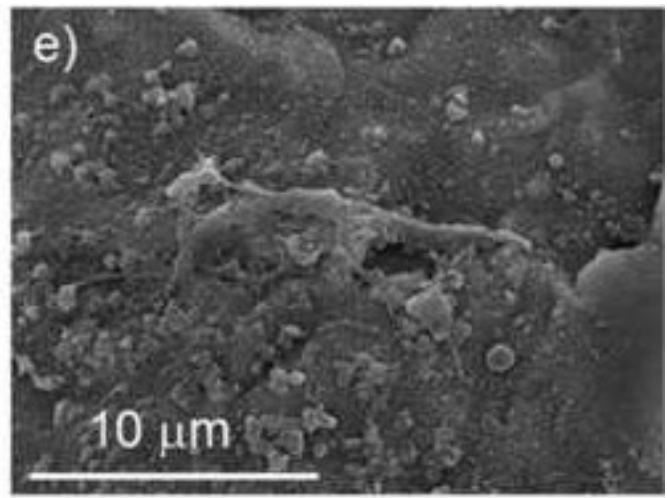

Figure 5

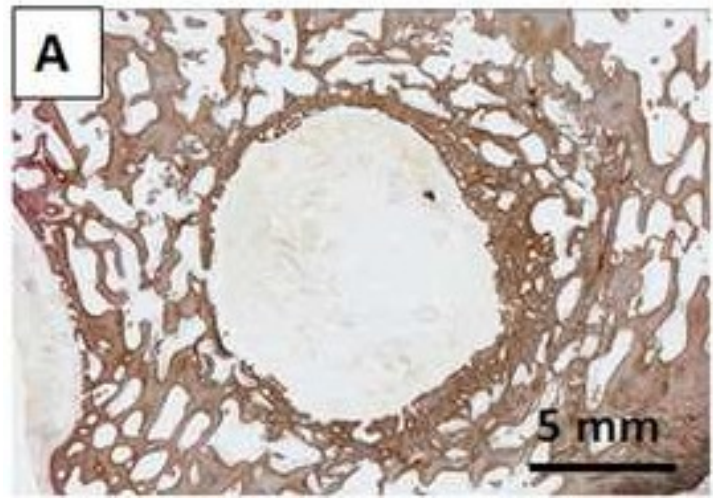
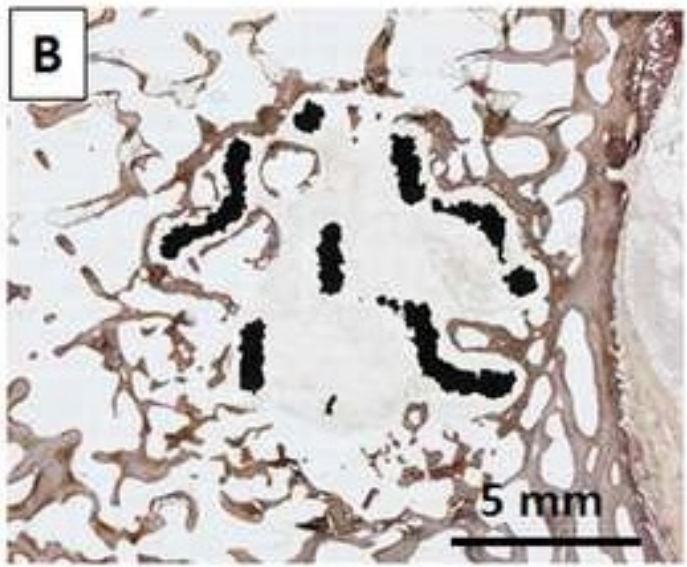
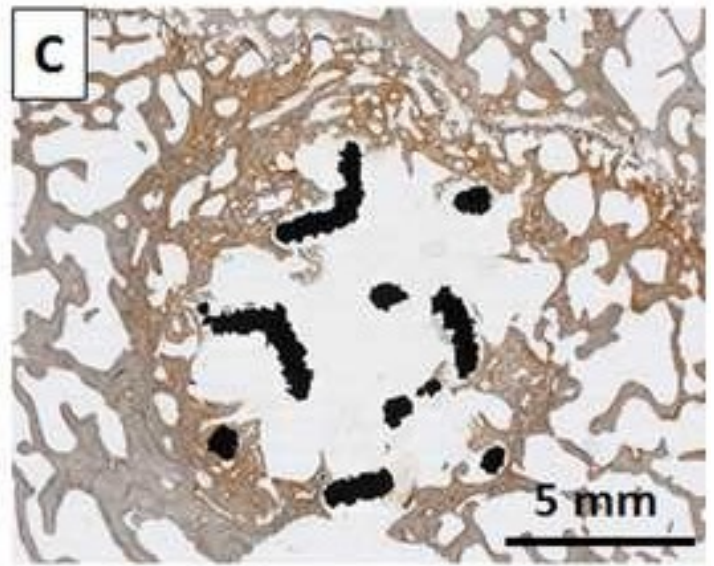
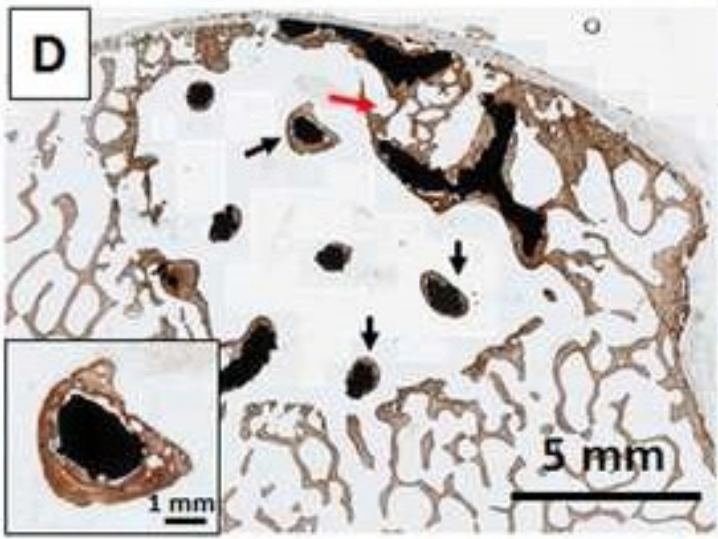
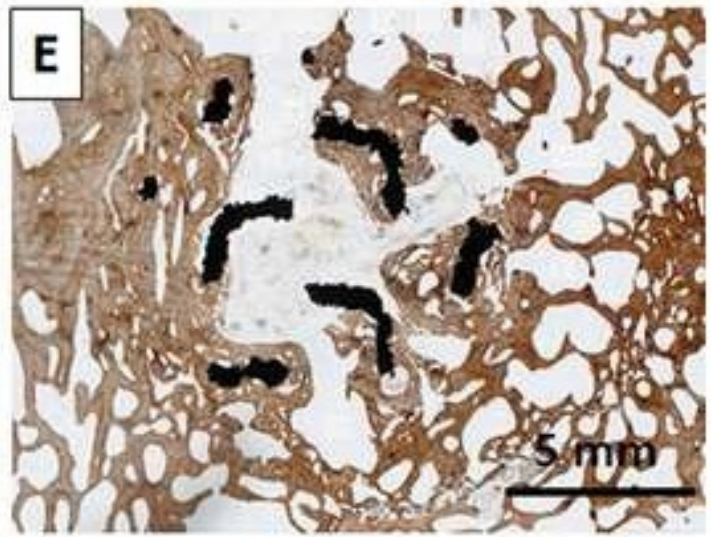

**Figures**
Click here to download high resolution image

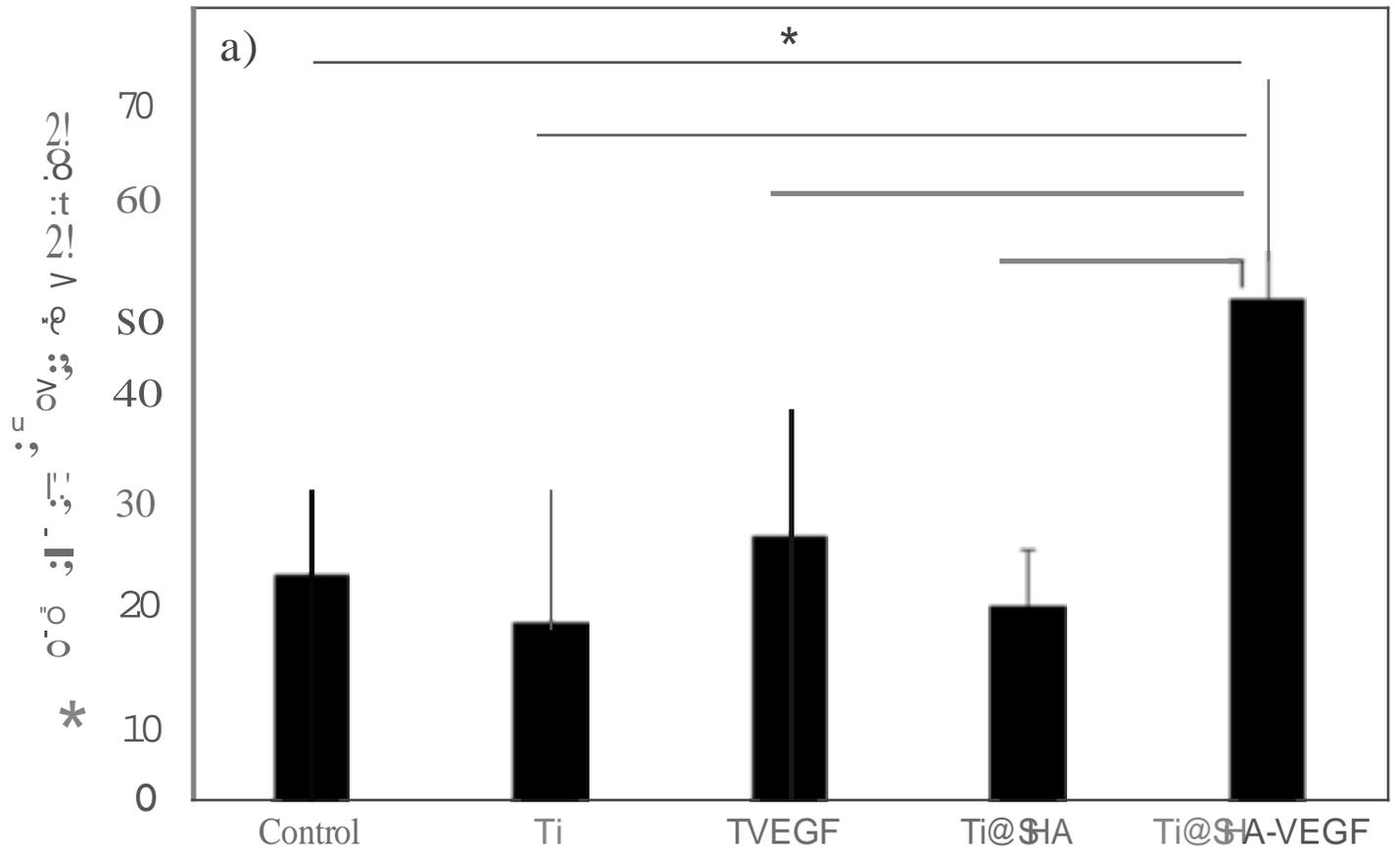

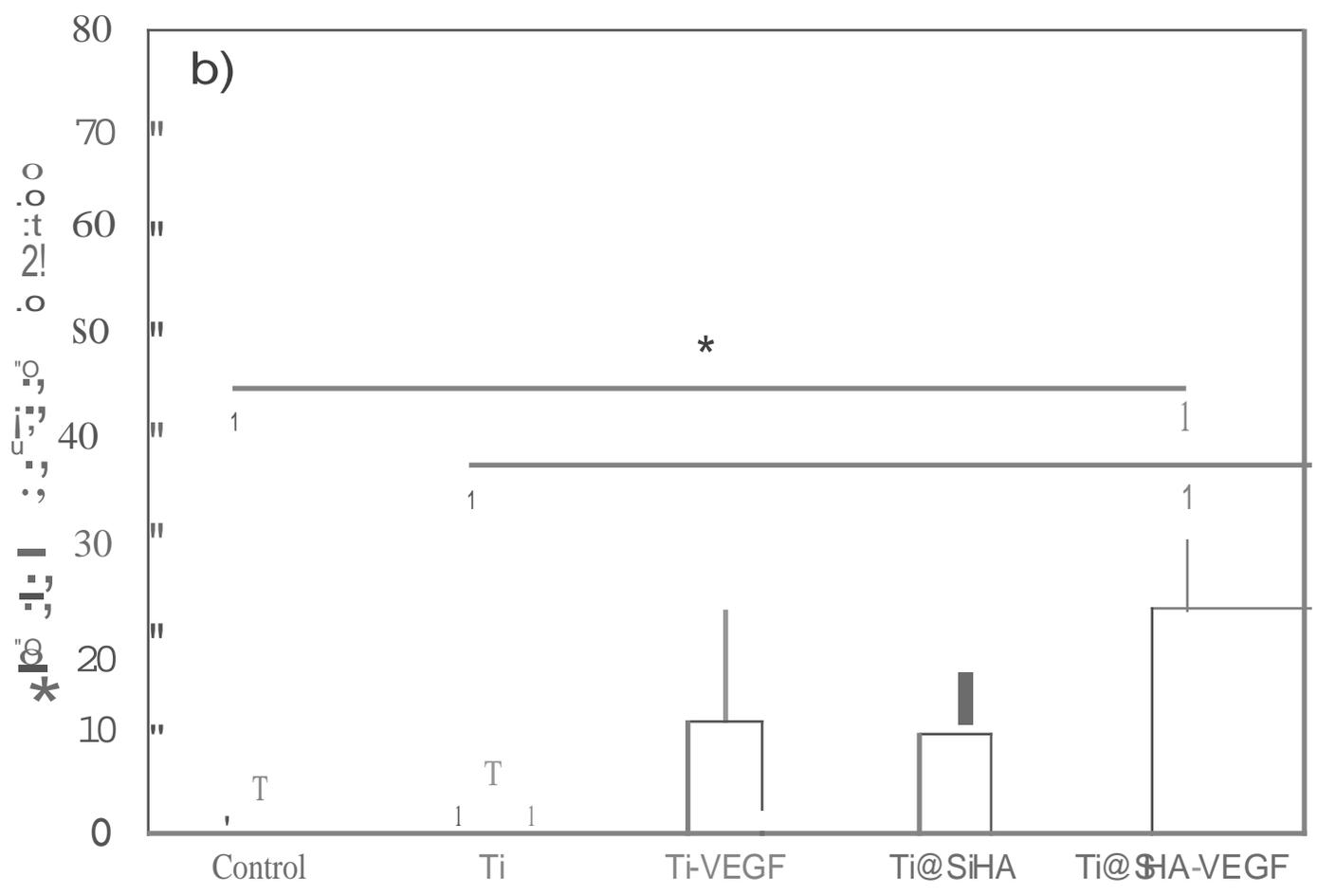

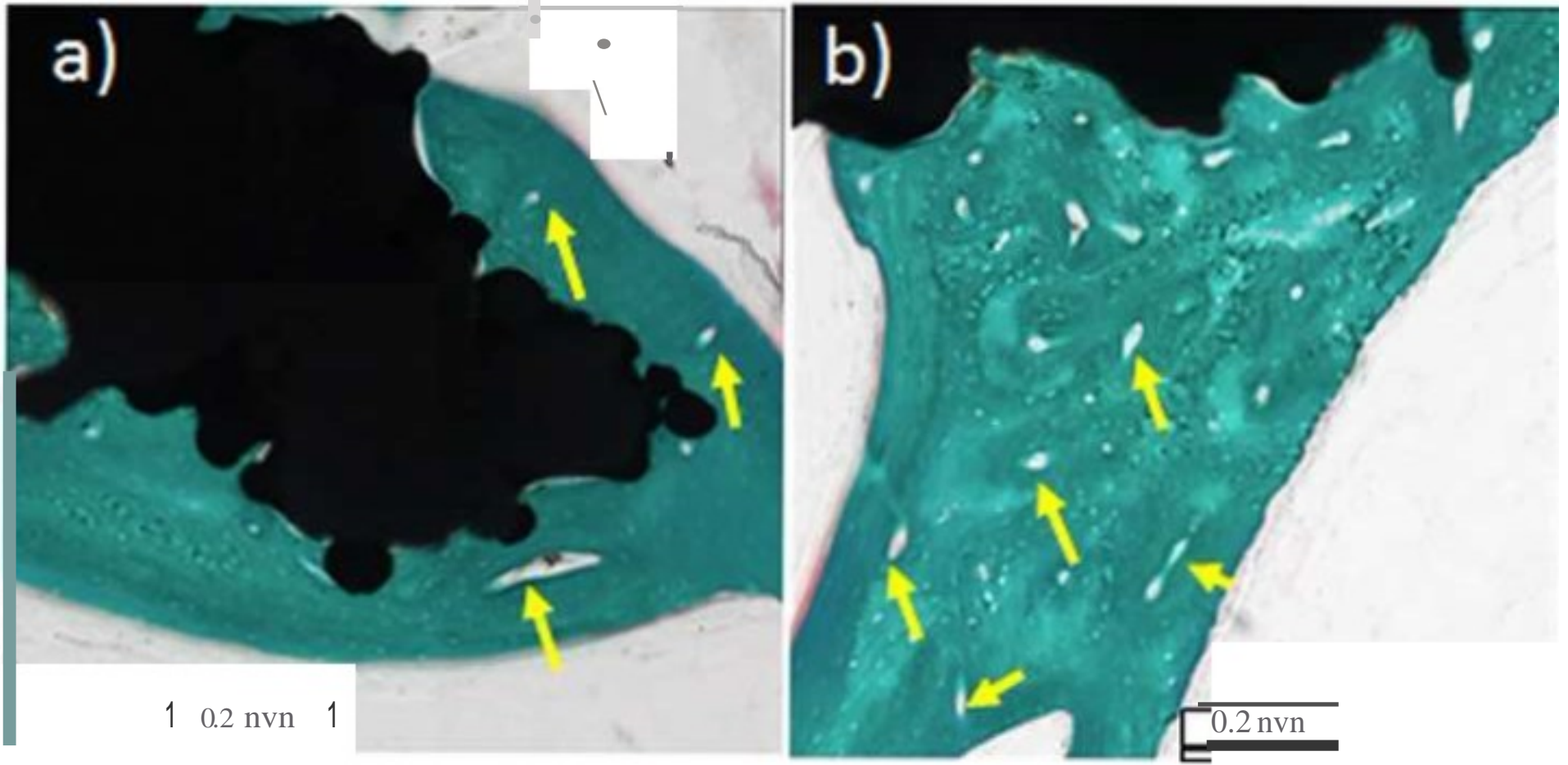

Figures

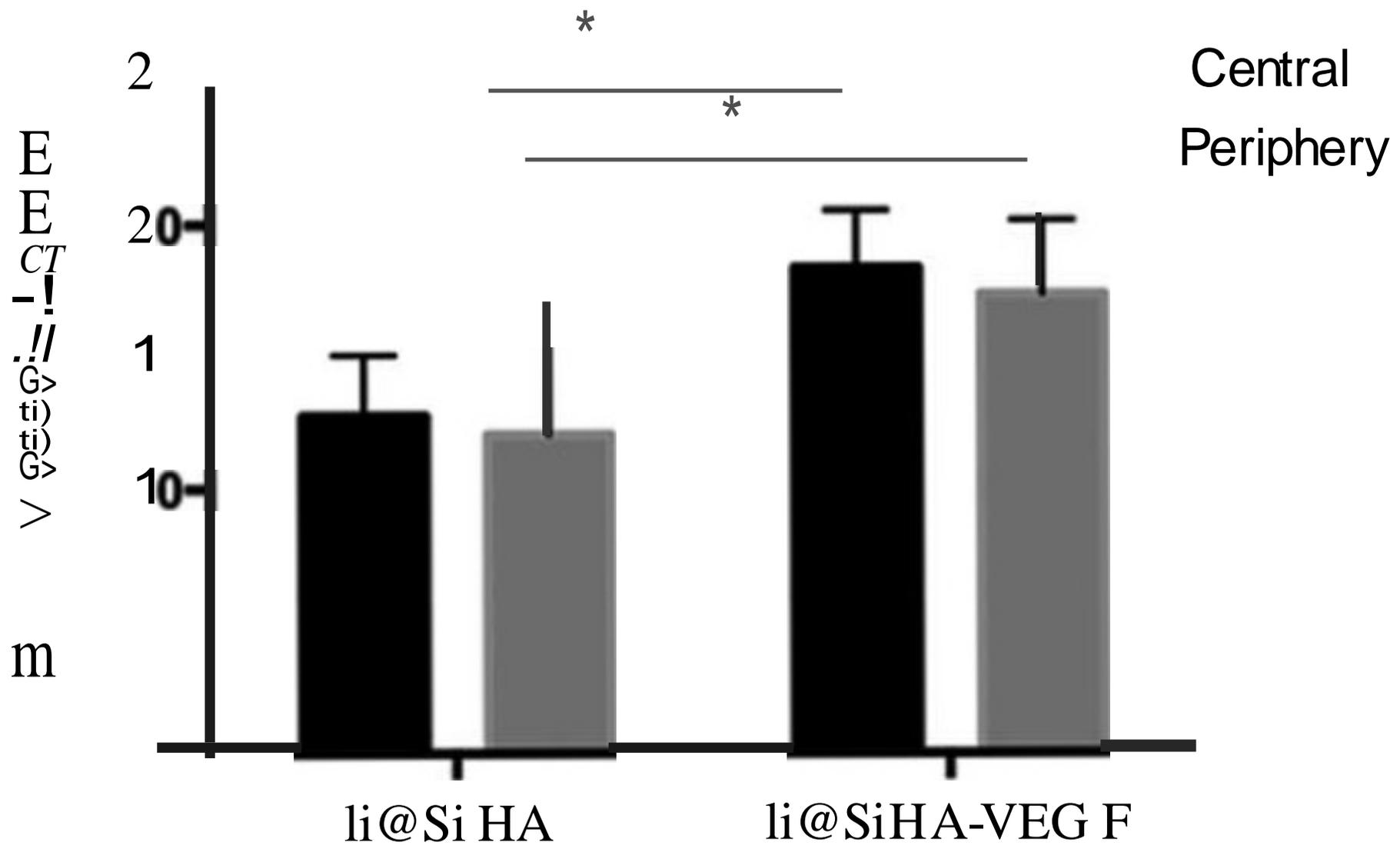

**Supplementary Material**

[Click here to download Supplementary Material: Supplementary information.docx](#)